\documentclass[preprint]{aastex}
\usepackage{emulateapj5}
\usepackage{onecolfloat5}
\usepackage{apjfonts}
\usepackage{epsfig}

\def\plotone#1{\centering \leavevmode
\epsfxsize= 1.0\columnwidth \epsfbox{#1}}

\def\gsim{\;\rlap{\lower 2.5pt
 \hbox{$\sim$}}\raise 1.5pt\hbox{$>$}\;}
\def\lsim{\;\rlap{\lower 2.5pt
   \hbox{$\sim$}}\raise 1.5pt\hbox{$<$}\;}
\newcommand{\be}{\begin{equation}}
\newcommand{\beq}{\begin{equation}}
\newcommand{\ba}{\begin{eqnarray}}
\newcommand{\ee}{\end{equation}}
\newcommand{\eeq}{\end{equation}}
\newcommand{\ea}{\end{eqnarray}}

\begin{document}
\twocolumn[
\submitted{Submitted to ApJ}
\title{Modeling the Counts of Faint Radio Loud Quasars: Constraints on the Supermassive Black Hole Population and Predictions for High Redshift}
\author{Zolt\'an Haiman,}
\affil{Department of Astronomy, Columbia University, 550 West 120th Street, New York, NY 10027, USA; zoltan@astro.columbia.edu}
\

\vspace{-0.5\baselineskip}
\author{Eliot Quataert, and Geoffrey C. Bower}
\affil{Department of Astronomy, 601 Campbell Hall, University of California at Berkeley, Berkeley, CA 94720, USA; eliot@astron.berkeley.edu; gbower@astron.berkeley.edu}
               
\begin{abstract}
We use a physically motivated semi--analytic model, based on the mass
function of dark matter halos, to predict the number of radio--loud
quasars as a function of redshift and luminosity.  Simple models in
which the central BH mass scales with the velocity dispersion of its
host halo as $M_{\rm bh}\propto \sigma_{\rm halo}^5$ have been
previously found to be consistent with a number of observations,
including the optical and X--ray quasar luminosity functions.  We find
that similar models, when augmented with an empirical prescription for
radio emission, overpredict the number of faint ($\sim 10\mu$Jy) radio
sources by 1--2 orders of magnitude. This translates into a more
stringent constraint on the low--mass end of the quasar black hole
mass function than is available from the Hubble and Chandra Deep
Fields. We interpret this discrepancy as evidence that black holes
with masses $\lsim 10^7~{\rm M_\odot}$ are either rare or are not as
radio-loud as their more massive counterparts.  Models that exclude
BHs with masses below $10^7~{\rm M_\odot}$ are in agreement with the
deepest existing radio observations, but still produce a significant
tail of high--redshift objects.  In the 1-10GHz bands, at the
sensitivity of $\sim 10\mu$Jy, we find surface densities of $\sim
100$, $\sim 10$, and $\sim 0.3$ deg$^{-2}$ for sources located at
$z>6$, $10$, and $15$, respectively.  The discovery of these sources
with instruments such as the {\em Allen Telescope Array (ATA)}, {\em
Extended Very Large Array (EVLA)}, and the {\em Square Kilometer Array
(SKA)} would open a new window for the study of supermassive BHs at
high redshift.  We also find surface densities of $\sim 0.1$
deg$^{-2}$ at $z > 6$ for mJy sources that can be used to study 21 cm
absorption from the epoch of reionization.  We suggest that, although
not yet optically identified, the FIRST survey may have already
detected $\sim 10^3-10^4$ such sources. \\ \\
\end{abstract}]


\section{Introduction}
\label{sec:intro}

The past few years have seen significant progress in probing the
ultra--high redshift universe, with both galaxies (e.g. Spinrad et
al. 1998; Hu et al. 2002; Rhoads et al. 2003; Kodaira et al. 2003) and
quasars (Fan et al. 2000, 2001, 2003) being discovered in increasing
numbers around and beyond redshift $z=6$ (see recent reviews by
Spinrad 2003 and Taniguchi 2003).  In hierarchical structure formation
scenarios in cold dark matter (CDM) cosmologies, the first baryonic
objects appear at still higher redshifts: at $z\approx 20-30$, when
the first high--$\sigma$ peaks collapse near the Jeans scale of $\sim
10^5~{\rm M_\odot}$ (Haiman, Thoul \& Loeb 1996; see Barkana \& Loeb
2001 for a recent review).  Radiative cooling is efficient in the
dense gas that has collapsed on these scales, and in principle, it can
facilitate efficient formation of stars and black holes (BHs).
Indeed, significant activity must have taken place at high redshifts,
in order to reionize the intergalactic medium (IGM) by $z\sim 15$
(Spergel et al. 2003).

The deepest detections of galaxies and quasars to date have been
obtained at optical or near infrared (NIR) wavelengths, where the
objects were identified in broad--band filters by their continuum, or
in narrow--band imaging observations by their Lyman--$\alpha$
emission.  The expected number of faint sources in future, deep NIR
observations have been studied extensively in the context of
hierarchical structure formation, using simple semi--analytic models.
Haiman \& Loeb (1997; 1998) showed that if halos collapsing at high
redshifts have reasonable star or BH formation efficiencies, they can
be detected in the NIR continuum in great numbers, with surface
densities possibly reaching $\sim1000$ sources per arcmin$^{-2}$ out
to redshifts $z \gsim 10$.  Similar models predict that a few $z>6$
quasars per $\sim 100$ arcmin$^{-2}$ could be visible in soft X--ray
bands at the flux limits already accessible to deep ${\it Chandra}$
and ${\it XMM}$ observations (Haiman \& Loeb 1999; Wyithe \& Loeb
2003).

Predictions analogous to those above, based on physically motivated
structure formation models, are currently lacking in the radio bands.
Observations of $z \lsim 6$ quasars have established that a
significant fraction ($\sim 10\%$) of these objects are bright in the
radio.  Although some of the detailed physics responsible for this
emission remains elusive, it is known to be a direct consequence of
outflows generated by accretion onto a central massive BH.  In this
scenario, one would expect that the population of radio loud quasars
extends to $z\gg 6$, to the epoch when the first BHs appeared and
started to accrete.  In fact, if the radio--loudness distribution does
not evolve strongly at high redshift (Ivezic et al. 2002; Petric et
al. 2003), and if high redshift supermassive BHs (SMBHs) radiate close
to their Eddington limit, then black holes with masses as small as
$M\sim 10^6{\rm M_\odot}$ should have radio flux densities that are
already being reached by the deepest existing observations ($\sim
10\mu$Jy at $\sim$ GHz frequencies).  {\it The purpose of this paper
is to confront simple models of the radio-loud quasar population with
current observations and to put forward predictions for the counts at
redshift and flux thresholds that are beyond the current observational
limits.} While such extrapolations are necessarily uncertain, the
detection of these objects would provide important constraints on the
formation and growth of the first SMBHs (see Haiman \& Quataert 2004
for a recent review).

Our predictions are especially timely in light of recent data on the
reionization history of the intergalactic medium (IGM).  On the one
hand, SDSS quasar spectra imply that we may have reached the neutral
epoch at $z\sim 7$ (e.g. Fan et al. 2002).  On the other hand,
observations of the cosmic microwave background (CMB) anisotropies
suggest that the universe was significantly ionized as early as $z\sim
15$ (Kogut et al. 2003; Spergel et al. 2003).  This behavior is a
challenge to reionization models. Pinning down the value of the
neutral fraction just beyond $z\sim 6$ would be of great value in
elucidating these models (see, e.g. Haiman 2003 for a recent review).
One promising probe of neutral hydrogen at high redshift is to study
redshifted 21cm absorption and emission features.  Background density
fluctuations from the ``21cm forest'', both in absorption (Carilli et
al. 2002) and in emission (Madau et al. 1997); absorption (Furlanetto
\& Loeb 2002) or emission (Iliev et al. 2002) from neutral gas in
discrete minihalos; and a sharp step--feature analogous to the
Gunn-Peterson trough (Shaver et al. 1999) could all, in principle, be
detected against a bright enough background source.  Such observations
would provide a powerful probe of the amount and distribution of
neutral hydrogen in the high--redshift IGM.  These studies will
obviously depend critically on the number of available radio sources
(although 21 cm features can also be studied against the CMB;
e.g., Tozzi et al. 2000).

The rest of this paper is organized as follows.  In
\S~\ref{sec:quasars} we summarize the phenomenology of radio--loud
quasars, and in \S~\ref{sec:model}, we briefly discuss the model we
adopt to describe their abundance and evolution.  In
\S~\ref{sec:BHmass}, we compare this model to existing observations
and in \S~\ref{sec:predict} we present our predictions for even higher
redshift.  In \S~\ref{sec:discuss} we conclude with a discussion of
our results and their implications.  Throughout this paper, we adopt
the background cosmological parameters as measured by the {\it WMAP}
experiment, $\Omega_m=0.27$, $\Omega_{\Lambda}=0.73$,
$\Omega_b=0.044$, $h=0.71$, $\sigma_{8h^{-1}}=0.9$ and $n_s=1$
(Spergel et al. 2003).

\section{Radio-Loud Quasars}
\label{sec:quasars}

Radio emission from a relativistic outflow is a ubiquitous feature of
accretion onto a central BH, from X-ray binaries (e.g., Fender 2001)
and Seyferts (e.g., Ho \& Peng 2001), to radio galaxies and quasars
(e.g., Urry \& Padovani 1995). In the context of active galactic
nuclei, the emission can include both an unresolved component near the
nucleus (e.g., the ``core'') and a spatially extended component such
as radio lobes.  The former is probably due to dissipation in the jet
(by, e.g., internal shocks or MHD instabilities) while the latter is
due to the interaction of the jet with the ISM or IGM (e.g., Begelman,
Blandford, \& Rees 1984).  Since jet production is determined by local
physics near the BH -- e.g., via a collimated wind originating in the
disk or via the Blandford-Znajek mechanism -- it is reasonable to
expect that jets will be launched from very high redshift BHs as well.
Moreover, although the extended emission from jets might be expected
to evolve with redshift as the conditions in the IGM change, the
nuclear emission is determined by relatively local physics and so is
likely to be much less sensitive to the ambient environment around the
BH.  Indeed there is already suggestive evidence that the fraction of
radio-loud quasars does not evolve significantly even out to $z
\approx 6$ (Ivezic et al. 2002; Petric et al. 2003).

\section{Modeling the High Redshift Population}
\label{sec:model}

Our semi--analytical approach is a simplified version of the
Monte--Carlo merger--tree models for the evolution of the AGN
population found in the literature of hierarchical galaxy formation
(Kauffmann \& Haehnelt 2000; Menou et al. 2001; Volonteri et
al. 2003). Its main ingredients are (1) the mass function of dark
matter halos; (2) the ratio $M_{\rm bh}/M_{\rm halo}$ of black hole to
halo mass as a function of $M_{\rm halo}$ and redshift $z$; (3) the
probability distribution of radio loudness (defined here as the ratio
of the radio to optical flux density); and (4) the duty cycle of
quasar activity.  Note that our simple model is not applicable at
$z\lsim 2$, where a single dark matter halo may host more than one
quasar.

{\em (1) Halo mass function.} We assume that SMBHs populate dark
matter halos, whose abundance $dN/dM_{\rm halo}(M_{\rm halo},z)$
follows the form derived from cosmological simulations (Jenkins et
al. 2001, equation 9). The cosmological power spectrum is computed
from the fitting formulae of Eisenstein \& Hu (1999).

{\em (2) Black Hole Mass.} We then assume that each halo harbors a 
central massive black hole of mass
\beq
M_{\rm bh}= 10^6 
\left(\frac{M_{\rm halo}}{1.5\times 10^{12}{\rm M_\odot}}\right)^{5/3} 
\left(\frac{h}{0.71}\right)^{5/3} 
\left( 1+z \right)^{5/2} {\rm M_\odot}.
\label{eq:Mbh}
\eeq
The scaling with $M_{\rm halo}$ and $z$ in this equation is equivalent
to $M_{\rm bh}\propto \sigma_{\rm halo}^5$, where $\sigma_{\rm halo}$
is the velocity dispersion of the dark matter halo.\footnote{There are
additional cosmology-- and redshift--dependent terms in the standard
relation between $M_{\rm halo}$ and $\sigma_{\rm halo}$ obtained from
the virial theorem (e.g. Iliev \& Shapiro 2001), but these approach a
constant value at the high redshifts considered here, $z\gsim 3$, and
can be absorbed into the normalization of equation~(\ref{eq:Mbh}).}
This scaling is consistent with the locally measured relation between
the central velocity dispersion $\sigma_c$ and BH mass in nearby
galaxies, when the conversion between $\sigma_c$ and $\sigma_{\rm
halo}$ is taken into account using models for the DM halo profile
(Ferrarese 2002, equation 6). Finally, equation~(\ref{eq:Mbh}) is also
consistent with a physical picture in which feedback from the quasar's
radiation and outflows determines the size of the black hole (by
shutting off accretion once its cumulative energy output has reached
the binding energy of the accreting gas; Silk \& Rees 1998; Haehnelt
et al. 1998; Wyithe \& Loeb 2003).  We chose the normalization in
equation (1) by requiring the model to predict the luminosity function
of optical quasars at $z\gsim 3$ for the duty cycle of $2\times10^7$
years (see Haiman \& Loeb 1998 for more details of the method).  The
normalization we obtain is very close to the value found in a recent,
more elaborate model by Wyithe \& Loeb (2003, equation 4) and implies
that $\sim 10\%$ of the quasar's energy output is deposited in the
surrounding gas.

{\em (3) Radio Loudness Distribution.}  The theory of radio emission
from jets is not sufficiently well--understood to make reliable
predictions for the radio flux of an accreting BH.  We instead follow
an empirical approach. We assume that each quasar shines at the
Eddington luminosity for a timescale $t_q$ (discussed below). We then
compute the BH's flux $F_i$ in the $i$--band using the template
spectrum of Elvis et al. (1994).  To make predictions in the radio, we
use an observationally determined radio loudness distribution.  Ivezic
et al. (2002) compare the SDSS and FIRST surveys to infer the fraction
of quasars with a given radio loudness $R$, where $R \equiv
\log_{10}[F_{1.4}/F_i]$ is the 1.4 GHz radio flux density relative to
the $i$--band optical flux density.  An approximate fit to their
results (Fig. 19) gives
\beq 
N(R) = 0.5 f_l \exp[-(R-{\bar R})^2/\sigma^2] \label{RL}
\label{eq:NR}
\eeq 
where $f_l \approx 10\%$ is the fraction of quasars that are radio
loud, $\sigma \approx 2/\sqrt\pi$, and ${\bar R} \approx 2.8$ is the
average radio-loudness.  Note that the ratio $R$ defined by Ivezic et
al. (2002) is for the {\it observed} 1.4 GHz and $i$--band ($\sim
8000$\AA) flux densities.  However, the radio-loudness distribution
described by equation~(\ref{RL}) must physically arise between the
emitted rest--frame luminosities of the sources. The mean redshift of
the sources used to derive equation (2) is $\langle z \rangle\sim 1$
(see Fig. 21 in Ivezic et al. 2002); we therefore assume that $R
\equiv \log_{10}[L_{2.8}/L_{i/2}]$, where $L_{2.8}$ and $L_{i/2}$ are
the rest--frame luminosities at $1.4(1+\langle z\rangle)=2.8$ GHz and
at $8000/(1+\langle z\rangle)=4000$\AA, respectively.  Note also that
equation~(\ref{RL}) has been normalized to $f_l$. The remaining $90
\%$ of radio-quiet quasars also produce radio emission, but the fluxes
are too small to be of interest here.  As mentioned above, there is no
evidence for significant evolution in the radio loudness distribution
with redshift (Ivezic et al. 2002; Petric et al. 2003), providing some
support for extrapolating the locally--determined distribution to yet
higher redshifts. Finally, for most of our calculations we assume that
the radio spectrum is flat, i.e., $\alpha = 0$ where $F_\nu \propto
\nu^{-\alpha}$.  However, we also present results for a steeper
spectrum, $\alpha = 0.5$, to illustrate the dependence on $\alpha$.

{\em (4) Duty cycle.} We assume a fixed lifetime of $t_q=2\times
10^7$yr for the cumulative duration of the radio--loud phase(s) for
each black hole.  This value is consistent with the time--scale for
Eddington limited accretion (Salpeter time), as well as with the
quasar lifetimes obtained from other independent arguments (see
Martini 2003 for a review, and Haiman \& Loeb 1998 and Haehnelt et
al. 1998 for discussions of how the lifetime can be uniquely related
to the scalings assumed in equation~[\ref{eq:Mbh}] above).  We
approximate the duty cycle of activity, defined as the fraction of
quasars that are active at a given time, as $f_{\rm duty}=t_q/t_H(z)$,
where $t_H(z)$ is the age of the universe at redshift $z$. This simple
assumption is justified by noting that the timescale for the formation
of new BHs is approximately $t_{\rm form}\sim (dN/dM)/(d\dot{N}/dM)
\sim t_H(z)$.

It is important to note that both $t_q$ and $t_{\rm form}$ can, in
general, be a function of both redshift and BH (or halo) mass.  In
particular, in extended Press-Schechter models, the typical halo age
is a decreasing function of $t_H(z)$ at high redshift and high halo
masses.  For example, one may define the duty cycle as the fraction of
all halos younger than $t_q$, with the halo age--distribution taken as
the distribution of the half--mass assembly times in the extended
Press--Schechter formalism (e.g., equation 2.26 in Lacey \& Cole
1993).  We find that using this definition would increase the number
counts we predict at high redshift ($10\lsim z \lsim 15$) by factors
of $\approx 3-6$ (in Figures~\ref{fig:rlcounts}
and~\ref{fig:rlcounts2} below).  On the other hand, if $t_q$ were
related to the dynamical time in the halo, rather than to the Salpeter
time (as proposed in Wyithe \& Loeb 2003), $t_q\propto (1+z)^{-3/2}$,
which would nearly cancel this increase in counts at high redshift.

\begin{figure}[t]
\plotone{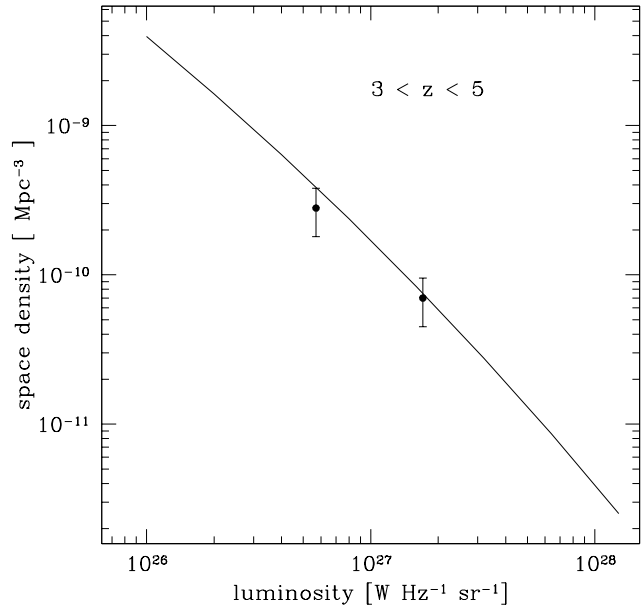}
\caption{Predicted luminosity function of radio--loud quasars in the
redshift bin $3<z<5$.  The space density of all quasars brighter than
a fixed luminosity is plotted against the luminosity at 5 GHz.  The
data points are taken from Hook, Shaver \& McMahon 1998 (their Figure
4).}
\label{fig:checkf} 
\end{figure}

\begin{figure}[t]
\plotone{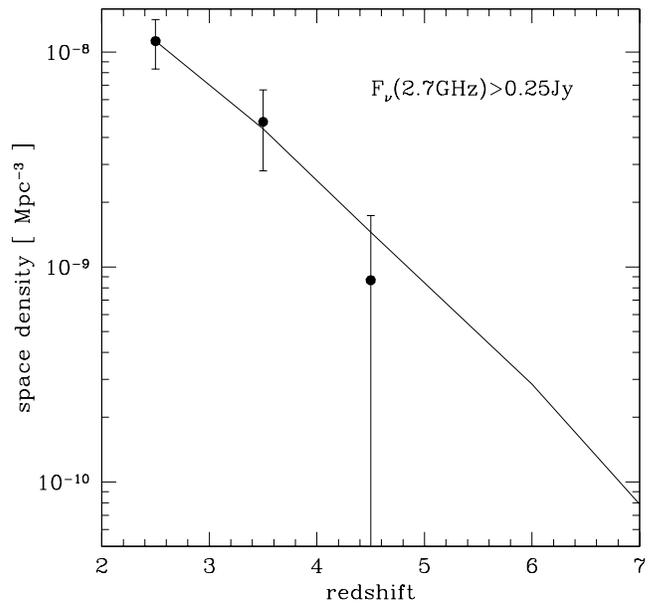}
\caption{Predicted space density of radio--loud quasars as a function
of redshift.  A 2.7 GHz flux density limit of 0.25 Jy was assumed. The
data-points with error bars are taken from Hook, Shaver \& McMahon
1998 (their Figure 6).}
\label{fig:checkz} 
\end{figure}

Given the above assumptions, the number of radio loud quasars per unit
redshift and solid angle, with a flux density brighter than $F_{1.4}$
at redshift $z$, is given by
\beq
\frac{dN}{dzd\Omega}\left(F_{1.4},z\right) =
\frac{dV}{dzd\Omega}\frac{t_q}{t_H(z)}
\int_0^\infty dM f(>R[M,F_{1.4}]) \frac{dN}{dM},
\label{eq:dNdzdom}
\eeq 
where $M$ is the halo mass, $dV/dzd\Omega$ is the cosmological volume
element, $t_H(z)$ is the age of the universe at redshift $z$, and
$f(>R)=\int_R^\infty dR N(R)$ is the fraction of sources with radio
loudness $R$ or higher.  Inside the integral, $R=R(M,F_{1.4})$ is
obtained by requiring that the BH residing in the halo of mass $M$,
whose mass and optical flux are fixed by assumption, should have a
given radio flux density $F_{1.4}$.

We have verified that the above model is consistent with the optical
and X--ray quasar luminosity functions at $z\gsim 3$ (following Haiman
\& Loeb 1998; 1999).  In addition, we have computed the number of
high--redshift sources that should be detectable in the {\it Chandra}
Deep Field North.  Adopting the spectral template of Elvis et
al. (1994), a $\sim 10^8 M_\odot$ BH with Eddington luminosity at
redshift $z=11$ will have a flux of $2\times10^{-16}~{\rm
erg~s^{-1}~cm^{-2}}$ in the soft X--ray band (see Fig.1 in Haiman \&
Loeb 1999).  Using the same assumptions described above, we find that
our model predicts $\sim 6$ quasars at $z>3$ and $\sim 1$ quasar at
$z>5$; these numbers are consistent with the observed number of faint
X--ray sources (Alexander et al. 2003; Barger et al. 2003).

In order to ensure further that our model is consistent with existing
observations of radio--loud quasars, in Figures~\ref{fig:checkf} and
\ref{fig:checkz} we show the results of our model for the luminosity
function and its redshift evolution at the bright flux densities
($\sim 0.2$Jy) at which it has been determined at $z\gsim 2$.  In
Figure~\ref{fig:checkf}, we show a luminosity range in the highest
redshift bin $3<z<5$ that is used in Figure 4 of Hook, Shaver \&
McMahon (1998, reproduced by the points with error bars).  In
Figure~\ref{fig:checkz} we show the redshift evolution at a 2.7 GHz
flux density limit of 0.25 Jy, so that our numbers can be directly
compared to the data in Figure 6 of Hook et al (reproduced by the
points with error bars).  As these two figures show, our simple model
is in good agreement with the available data on the high-redshift
radio LF.

\section{A Constraint on the SMBH Population from Existing Source Counts}
\label{sec:BHmass}

\begin{figure}[t]
\plotone{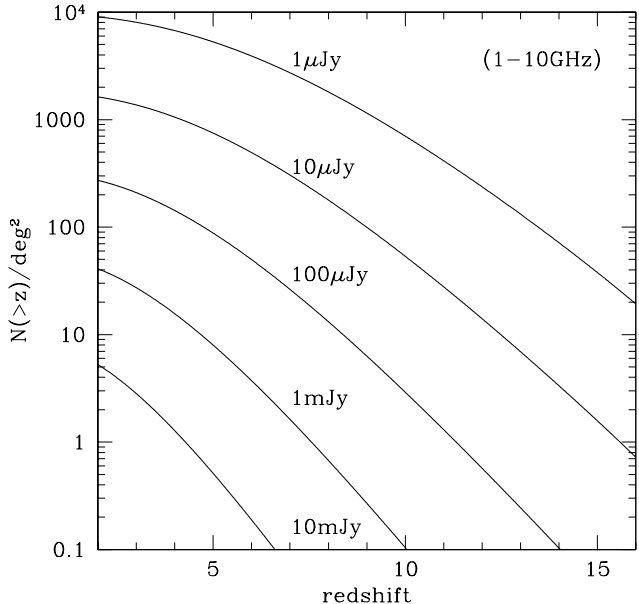}
\caption{Predicted number of radio--loud quasars as a function of
redshift (the counts are cumulative).  A flat spectrum, $\alpha = 0$,
is assumed, so the counts are the same at any frequency from $\approx
1-10$ GHz.  Figure 6 shows the dependence on spectral index $\alpha$.}
\label{fig:rlcounts} 
\end{figure}

Figure~\ref{fig:rlcounts} shows the predicted number counts in the
radio as a function of redshift for five different choices of the
threshold flux density.  For this figure we took $\alpha = 0$, i.e.,
$F_\nu \approx {\rm constant}$ (see Fig. 6 for results with $\alpha =
0.5$).  The same number counts are thus predicted for any frequency in
the $\approx 1-10$ GHz range; at sufficiently high frequencies ($\gsim
10-100$ GHz, depending on $z$), rest frame dust emission could become
important and dominate over jet emission.

\begin{figure}[t]
\plotone{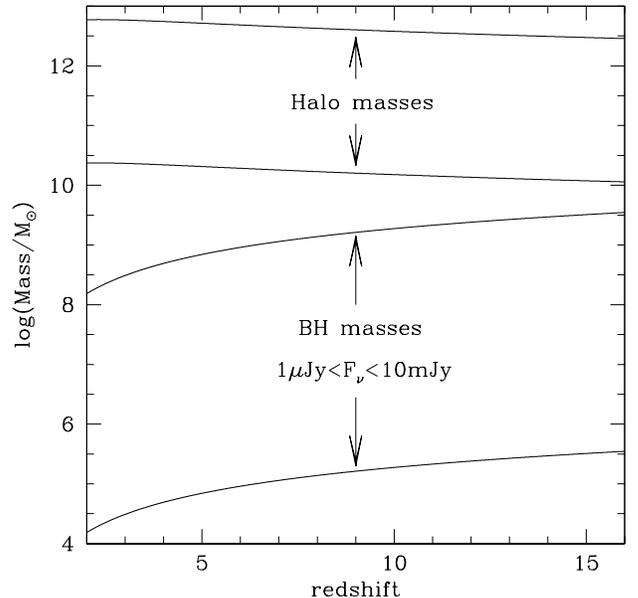}
\caption{The mass of the BH (lower pair of curves) and halo (upper
pair of curves) powering the typical radio--loud quasar, as a function
of redshift, for the range of flux densities used in
Figure~\ref{fig:rlcounts}.}
\label{fig:rlbhmasses} 
\end{figure}

Figure~\ref{fig:rlbhmasses} shows the mass of the typical BH at a
given flux density and redshift (i.e., a BH with the mean
radio--loudness ratio $R={\bar R}=2.8$).  The upper pair of curves
show the corresponding halo masses (whose range is reasonable).  The
important point revealed by this figure is that deep radio
observations at flux densities $\lsim 10\mu$Jy can probe the
high-redshift population of SMBHs at $M_{\rm bh}\sim10^4-10^6~{\rm
M_\odot}$, a range of masses that is not currently detectable in the
optical/X--ray bands.  In fact, it is not clear whether BHs as small
as $\lsim 10^{4-6}{\rm M_\odot}$ exist at the centers of galaxies.
Haiman, Madau \& Loeb (1999) obtained an empirical lower limit of
$\sim 10^6 M_\odot$ from the flattening of the optical quasar LF
required to avoid over-predicting the number of faint, high--redshift
quasars in the {\it Hubble} Deep Field.  Moreover, the smallest
directly measured mass for a central SMBH is just above $10^6~{\rm
M_\odot}$ (e.g. Gebhardt et al. 2000; Merritt \& Ferrarese 2001).  If
some holes as small as $10^4 (10^5)~{\rm M_\odot}$ do exist, our
results suggest that they could have radio flux densities $\sim 1 (10)
\ \mu$Jy at $z = 4$ and $\sim 0.1 (1) \ \mu$Jy at $z = 8$, and so
could be detected in very deep radio observations.

The results shown in Figure~\ref{fig:rlcounts} for the expected number
counts can, in fact, be directly compared to existing observations.
The FIRST survey (Becker et al. 1995) is using the Very Large Array
(VLA) to map $10,000$ deg$^{2}$ of the sky down to a point--source
sensitivity of $\sim 1$mJy. The survey is nearly complete, and at this
flux density threshold, it has revealed a surface density for discrete
sources of $\sim 75$ deg$^{-2}$, implying a total of $\sim 7.5\times
10^5$ sources over the entire survey area.\footnote{For the status and
results of the FIRST survey, see http://sundog.stsci.edu.}  In a
cross--correlation between SDSS and FIRST in a $1230$ deg$^{-2}$
region, Ivezic et al. (2002) identify $\sim 30\%$ of the FIRST sources
with SDSS sources; $\sim 17\%$ of this matched sample are quasars.
However, with the $N(R)$ distribution derived for the matched quasars
(see eq.~\ref{eq:NR}), the $1$mJy FIRST flux density threshold
corresponds to a typical magnitude $i=23.4$, well below the detection
threshold of SDSS ($i=21.5$). This suggests that many FIRST sources
must be quasars without SDSS counterparts. Ivezic et al. (2002;
Appendix B) estimate the fraction of FIRST sources that could be
quasars too faint for SDSS to detect.  Taking a conservative count of
optical quasars (from Pei 1995) down to a magnitude of $i=23.5$, and
combining it with the radio--loudness distribution $N(R)$, they find
that for every FIRST quasar detectable by SDSS ($i<21.5$), there
should be at least $\sim$ six quasars at that are below the SDSS
threshold.  This implies that, overall, the fraction of quasars in the
entire FIRST catalog should be $\sim 7\times 0.17\times 0.30=0.36$, or
$\sim 270,000$ quasars in total. This is in reasonable agreement with
the $\sim 400,000$ sources we predict at $z>2$ and $>1$mJy in
Figure~\ref{fig:rlcounts} (note that Fig.~\ref{fig:alpha} with $\alpha
= -0.5$ predicts a comparable number of such sources, $\sim 300,000$).

The fraction of quasars in the FIRST survey can be determined directly
by cross--correlating FIRST with optical surveys deeper than SDSS.  To
indicate the sensitivity of our results to model parameters, we note
that a factor of 4 reduction in the average radio loudness (to $\bar R
= 2.2$) of quasars near the FIRST threshold ($i \approx 23.5$) would
reduce our predicted number of $>1$ mJy quasars by a factor of
$\approx 4-5$ in the range $2<z<10$.  While there is no indication
that the radio loudness decreases for fainter quasars in the
SDSS-FIRST matched sample (to $i \approx 21.5$; see Ivezic et
al. 2002), such a decrease is not currently well constrained at the
faint end of the luminosity function that dominates our predicted mJy
counts in Fig. 3 (at $i \approx 23.5$, corresponding to BH masses of a
few $\times 10^7 M_\odot$).

We can also compare our model to recent radio observations at much
fainter flux levels.  One of the deepest radio images yet made is the
Fomalont et al. (2002) observation of the VLA field SA 13 (65
arcmin$^2$), which was observed to a depth of $7.5 \mu Jy$ (5
$\sigma$).  The calculations shown in Figure~\ref{fig:rlcounts}
predict $30$ radio-loud quasars in such an observation.  In contrast,
Fomalont et al. find a total of 34 radio sources, only two of which
are optically-detected quasars (10 \& 26 micro-Jy; see their Table 1).
In the remaining optically identified galaxies, there is no evidence
for quasar activity (though some could be lower luminosity
AGN).\footnote{It is important to stress that our radio-source
predictions assume that the bolometric luminosity of a BH is Eddington
(for a time $t_q$).  Thus the nuclear radio-sources would have optical
counterparts that dominate the light of the host galaxy.  In the
calculations presented here, we do not attempt to model
lower-luminosity AGN activity.  We would expect such sources to be
even more numerous than the bright quasars we focus on; it is thus
possible that some of the Fomalont et al. sources could be moderate
redshift lower-luminosity AGN.}  Finally, although nine of their
sources have no optical counterparts down to $I = 25.5$, and so in
principle could be very high redshift quasars, these sources have
steep radio spectra.  This suggests that they are starbursts and not
AGN (see also Richards 2000).  A similar conclusion is reached in the
observations of Richards et al. (1998), covering 65 arcmin$^2$ of the
HDF and surrounding fields at the sensitivity of 9$\mu$Jy (5
$\sigma$).  They find 29 sources in their statistically complete
sample, but none of these are quasars.  We are led to conclude that
the simple model in Figure~\ref{fig:rlcounts} over--predicts the
number of faint radio sources by about an order of magnitude.  Recall
that this model is consistent with the available optical and X--ray
data and the luminosity function and redshift evolution of bright
radio-loud quasars (Figs 1 \& 2).

This result implies that the radio LF has to flatten significantly at
high redshift at flux densities of $10-100\mu$Jy.  The simplest
interpretation of this flattening is that there is a characteristic BH
mass below which SMBHs either do not exist, are not accreting
significant gas, or are much less efficient at producing radio
emission.  To address this possibility, we computed the total number
of sources in a 65 arcmin$^2$ area down to the flux density threshold
of 7.5(9)$\mu$Jy.  We ignored the counts from all BHs with masses
below $M_{\rm crit}$ and varied $M_{\rm crit}$ until no more than 3
sources were predicted in either mock observation. We find that this
requires $M_{\rm crit}=10^7~{\rm M_\odot}$, or, equivalently, a
threshold velocity dispersion of $120~{\rm km~s^{-1}}$.  The resulting
modified number counts, ignoring all BHs with $M < M_{\rm crit}$, are
shown in Figure~\ref{fig:rlcounts2}.  As this figure reveals, ignoring
the low--mass BHs has little effect on the number of bright sources
($\gsim 1$ mJy).  Finally, an alternative way of stating this
constraint is that the average radio flux for BHs with $M \lsim 10^7
{\rm M_\odot}$ must decrease by a factor of $\gsim 15$ (i.e., from
$\bar R = 2.8$ to $\bar R \lsim 1.6$) in order to be consistent with
the deep $10\mu$Jy observations.

\section{Predictions for High--Redshift Counts for Future Instruments}
\label{sec:predict}

\begin{figure}[t]
\plotone{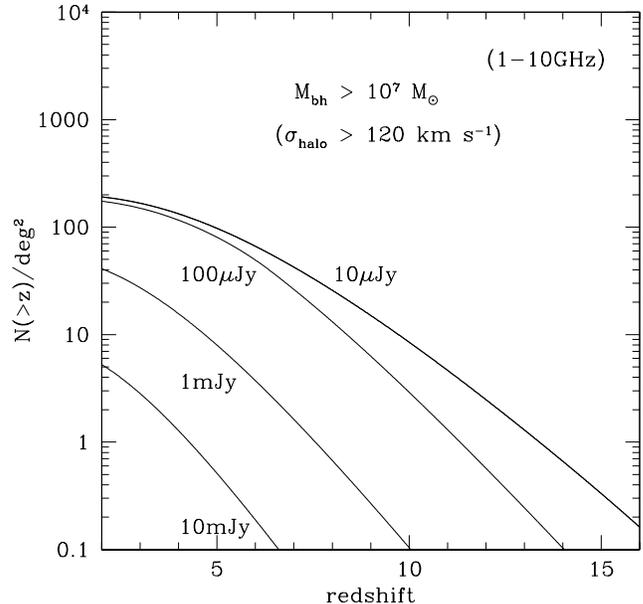}
\caption{Predicted number of radio--loud quasars as a function of
redshift with flux densities greater than the indicated levels,
ignoring the contribution of black holes with masses below $10^7~{\rm
M_\odot}$.}
\label{fig:rlcounts2} 
\end{figure}

Figure~\ref{fig:rlcounts2} extends the predictions of our model to
high redshift, for flux/counts combinations that could be probed by
future instruments.  For example, the Extended VLA (EVLA) will have a
sensitivity about 10 times better than that of the current
VLA.\footnote{See http://www.aoc.nrao.edu/evla} The Low Frequency
Array (LoFAr) will have $\sim$mJy sensitivity in the low frequency
(10--240 MHz) range.\footnote{See http://www.lofar.org} The Allen
Telescope Array (ATA), expected to be fully operational in 2006, has a
planned sensitivity similar to the VLA, but a significantly larger
field of view.\footnote{See http://www.seti.org/science/ata.html} It
can achieve a sensitivity of $7.5\mu$Jy at 1.4 GHz over $10$
deg$^2$ in about a week of observing.  The Square Kilometer Array
(SKA) will come online in about a decade.\footnote{See
http://www.skatelescope.org} Although its design is not yet final, SKA
will likely have $\sim100$ times the collecting area, but a $\sim 5$
times smaller field of view, than the ATA.  As
Figure~\ref{fig:rlcounts2} shows, imaging 10 deg$^2$ should
reveal a few faint sources out to redshifts as large as $z\sim 15$.

\section{Discussion}
\label{sec:discuss}

We have presented a simple physically motivated model for the SMBH
population and its evolution that fits the optical/IR and X-ray quasar
luminosity functions out to $z \approx 5$, and the luminosity function
and number counts of bright radio sources at high redshift.  This
model significantly overpredicts the counts at the $\approx 10 \
\mu$Jy level in deep radio observations.  This discrepancy can be
reconciled by postulating the existence of a lower limit to the SMBH
mass, below which SMBHs are either rare or do not produce as much
radio emission as their more massive counterparts.  We find that this
lower limit is $\approx 10^7~{\rm M_\odot}$, a constraint that is
approximately an order of magnitude more stringent than that available
from either the {\it Hubble} or {\it Chandra} deep fields (e.g.,
Haiman et al. 1999).  This constraint is especially interesting since
there are several SMBHs in local galaxies with dynamically determined
masses $\lsim 10^7 M_\odot$ (e.g. the central BHs in M32 -- van der
Marel et al. 1998 -- and in the Milky Way -- Sch\"odel et al. 2002;
Ghez et al. 2003).  In addition, Filippenko \& Ho (2003) argued for a
$\sim 10^5 M_\odot$ BH for the Seyfert galaxy in the late type
(bulgeless) spiral NGC 4395, and Barth et al. (2004) reached a similar
conclusion for the dwarf Seyfert 1 Galaxy POX 52.

If indeed SMBHs with masses below $\approx 10^7 M_\odot$ are rare,
this could be because it is difficult to form SMBHs in shallow
potential wells (e.g., Haehnelt et al. 1998; Haiman et
al. 1999). Alternatively, during the coalescence of SMBH binaries, the
remnant BH receives a ``kick'' velocity of up to several hundreds of
km s$^{-1}$ due to the net linear momentum carried away by
gravitational waves (e.g., Favata et al. 2004).  This kick may be
sufficient to unbind lower-mass BHs from their host galaxies, leading
to a dearth of low-mass BHs in galactic nuclei (e.g., Madau \&
Quataert 2004; Merritt et al. 2004).

An alternative explanation for the discrepancy between our models and
the number counts of faint radio sources is that their bright
accretion phase (near Eddington) is significantly shorter than that of
their high--mass counterparts (note, however, that the Salpeter time
characterizing the growth of SMBHs is independent of BH mass), or that
lower mass BHs with $M \lsim 10^7 M_\odot$ are intrinsically less
radio loud (we find that a reduction by a factor of $\approx 15$, or
changing $\bar R = 2.8$ to $\bar R \lsim 1.6$, is needed). There are
indeed some suggestions in the literature that very massive BHs are
preferentially radio loud (e.g., Laor 2000; Lacy et
al. 2001). However, other analyses suggest instead that the strongest
correlation is between radio loudness and Eddington ratio (e.g., Ho
2002), consistent with the assumptions used here.

\begin{figure}[t]
\plotone{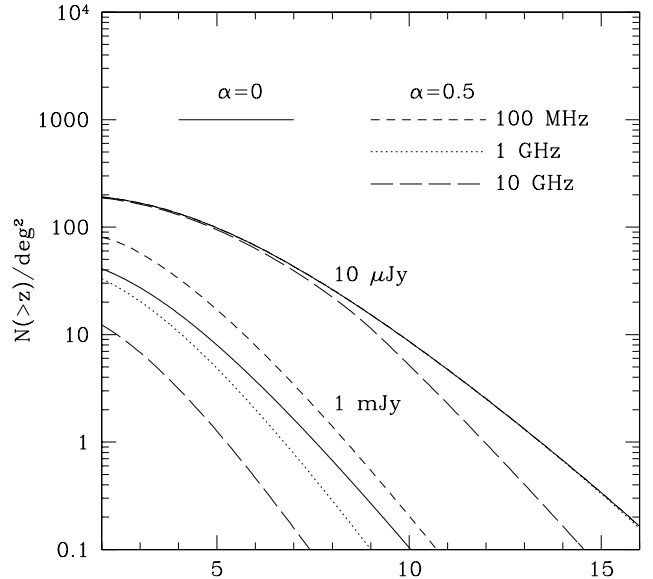}
\caption{This figure shows the sensitivity of our predictions for the
$10\mu$Jy and 1mJy counts to the assumed spectral slope $\alpha$ of
the radio spectrum $F_\nu\propto\nu^{-\alpha}$.  The counts for
$\alpha=0$ are independent of frequency.  When $\alpha=0.5$ is
assumed, however, the counts increase with decreasing frequency, as
shown by a comparison between the predictions at 100 MHz, 1GHz, and
10 GHz (short--dashed, dotted, and long--dashed curves,
respectively).}
\label{fig:alpha} 
\end{figure}

Our results also show that even in the presence of a low--mass cutoff
in the distribution of radio-emitting SMBHs (which we assume is
independent of redshift), a significant number of sources at redshifts
as high as $z\sim 15$ could be detectable.  In the 1-10GHz bands, at
the sensitivity of $\sim 10\mu$Jy, we find surface densities of $\sim
100$, $\sim 10$, and $\sim 0.3$ deg$^{-2}$ for sources located at
$z>6$, $10$, and $15$, respectively (Fig.~\ref{fig:rlcounts2}).  These
predictions for the cumulative counts are insensitive to our choice of
average radio loudness $\bar R$.  The reason is that a typical $10^7
M_\odot$ BH radiating at Eddington at $z = 6$ produces $\approx 100 \
\mu$Jy if $R \approx 2.8$ and $10 \ \mu$Jy if $R \approx 1.8$.  Thus
even if we decrease the average radio flux by an order of magnitude,
i.e., from $\bar R \approx 2.8$ to $\bar R \approx 1.8$, the total
surface density of the faintest sources shown in
Figure~\ref{fig:rlcounts2} does not change significantly (i.e.,
although the predicted fluxes for most of the sources decreases from
$100 \ \mu$Jy to $10 \ \mu$Jy, the total number of sources with $F_\nu
\gsim 10 \ \mu$Jy is relatively unchanged).  Figure~\ref{fig:alpha}
shows that, except at very high $z$, these predictions are also
insensitive to the choice of the spectral index $\alpha$.

A particularly interesting question is the redshift evolution of
bright radio sources, which can be used to study redshifted 21cm
absorption features and hence the reionization history of the
universe.  Carilli et al. (2002) find that a $\sim 6$mJy source (at a
few 100 MHz) is needed to achieve the S/N necessary for 21 cm
absorption studies.  For a flat spectrum ($\alpha = 0$), our results
(Figure~\ref{fig:rlcounts2}) imply that $\sim 2.5$ such sources should
be available in a 10 deg$^2$ field at $z=6-7$, with $\approx 2,000$
sources available over the full sky in the redshift range $8<z<12$.
Figure~\ref{fig:alpha} shows that the predicted counts are larger by a
factor of few if we instead assume a steeper spectrum with $\alpha =
0.5$.  It is also important note that since these bright mJy sources
are produced by massive BHs, their counts are independent of the
low-mass cutoff considered above.  They are, however, sensitive to an
evolution in the average radio--loudness $\bar R$ (in contrast to the
counts of the faint sources). For example, decreasing $\bar R$ from
${\bar R}=2.8$ to ${\bar R}=2.3$ beyond $z>6$ would decrease the
number of 6mJy sources at $6<z<7$ and $8<z<12$ by a factor of $4.6$,
and $6.3$, respectively.  Another important point to note is that our
fiducial results, obtained assuming $\bar R=2.8$, predict that the
FIRST survey may have already detected $\sim 10^3-10^4$ quasars at
$\sim 1$mJy from redshift $z\gsim 7$.  The identification of these
quasars is a challenge, but should, in principle, be feasible with
deep optical/IR observations.

The extrapolation of the radio-source population to high redshifts is
necessarily uncertain; nevertheless, the results presented here should
serve as useful order of magnitude estimates. We also note that our
assumption of $M_{\rm bh}\propto\sigma^5$ is somewhat conservative; a
shallower relation would imply that the typical BHs reside in
lower--mass halos (with shorter lifetimes; see Haiman \& Loeb 1998 and
Haehnelt et al. 1998); their abundance would then decrease less
rapidly at high redshifts.

An important issue in designing future surveys is the relative merits
of area vs. depth. The DM halo mass function has an approximate
power--law shape at low masses $dN/dM \propto M^{-2}$, and the
velocity dispersion scales as $\sigma\propto M^{1/3}$.  As a result,
if $M_{\rm bh}$ scales as $\sigma^\alpha$, and if the flux is
proportional to $M_{\rm bh}$, then the number of sources will go as
$F dn/dF \propto F^{-3/\alpha}$.  The number of detections scales
linearly with the solid angle $\Delta\Omega$, and with the observation
time as $t^{1.5/\alpha}$; area is therefore more important than depth
for the empirically determined slopes of $\alpha \approx 4$ (Gebhardt
et al. 2000; Merritt et al. 2001; Tremaine et al. 2002). In principle,
at bright flux limits, corresponding to BH masses where the halo mass
function turns over and drops exponentially, decreasing the flux
threshold would result in a larger yield of sources.  However,
Figure~\ref{fig:rlcounts2} reveals that even at the brightest fluxes
shown, the source counts increase linearly with the flux threshold,
and it would therefore be more advantageous to cover a larger area.

To conclude, we briefly discuss how to identify the high redshift
radio sources.  At low flux densities ($\lsim 30 \ \mu$Jy), starbursts
dominate over quasars in deep radio observations (e.g., Richards et
al. 1998).  There is also a contribution from moderate redshift
low--luminosity AGN.  AGN can be distinguished from starbursts by
their flatter spectra and variability.  Isolating the high redshift
sources, however, will require optical/IR followup.  Our typical $10 \
\mu$Jy source at $z \approx 6$ and $z \approx 10$ is powered by a
$\approx 10^7 M_\odot$ BH (without the cutoff described above it would
be $\lsim 10^6 M_\odot$, but we would then overpredict the number of
such sources).  With the Elvis et al. (1994) spectral template,
normalized to a bolometric Eddington luminosity, these sources would
have a flux density of $\sim 0.3 \mu$Jy at observed wavelengths of
$\sim 1-5\mu$m, or an AB magnitude of $\sim 25.5$ (see, e.g., figure 1
in Haiman \& Loeb 1998).  They should be detectable in moderate
integrations with the {\it Hubble} or {\it Spitzer Space
Telescopes}.\footnote{The {\it Spitzer Space Telescope} has a
point--source sensitivity of $\sim 3\mu$Jy in a 10-second exposure at
3.6$\mu$m, see http://ssc.spitzer.caltech.edu/irac/sens.html} The very
bright $\sim$mJy sources at $z \approx 6-10$, relevant for 21 cm
absorption studies, should have flux densities of $\sim 3\mu$Jy (or AB
magnitudes of $\sim 23$ at $\sim 1-5\mu$m), and should be detectable
in short exposures with {\it HST} or {\it Spitzer}, and potentially
from the ground as well. The discovery and confirmation of even a few
radio sources at $z>10$ by instruments such as the {\em Allen
Telescope Array (ATA)}, {\em Extended Very Large Array (EVLA)}, and
the {\em Square Kilometer Array (SKA)} would open a new window for
the study of supermassive BHs at high redshift, and of the
pre--reionization universe.  Of even more immediate interest is the
prediction from our models that, although not yet optically
identified, the FIRST survey may have already detected $\sim
10^3-10^4$ distant $z>7$ quasars.  Deep surveys, such as NOAO Deep
Wide-Field Survey (NDWFS), VIRMOS and DEEP may cover the area
necessary to identify a handful of these high-$z$ FIRST sources.

\acknowledgements{We thank David Helfand for useful discussions and a
careful reading of the manuscript.  ZH acknowledges financial support
by NSF through grants AST-03-07200 and AST-03-07291.  EQ acknowledges
financial support from NASA grant NAG5-12043, NSF grant AST-0206006,
an Alfred P. Sloan Fellowship, and the David and Lucile Packard
Foundation.}

\vspace{-2\baselineskip}

\end{document}